\documentclass{PoS}

\title{A new high sensitivity search for neutron-antineutron oscillations at the ESS}

\ShortTitle{Neutron-antineutron oscillations at the ESS}

\author{\speaker{David Milstead}\thanks{On behalf of the NNbar@ESS collaboration.}\\
        Stockholm University\\
        E-mail: \email{milstead@fysik.su.se}}

\abstract{A sensitive search for neutron-antineutron oscillations can provide a unique probe of some of the central questions in particle physics and cosmology: the energy scale and mechanism for baryon number violation, the origin of the baryon-antibaryon asymmetry of the universe, and the mechanism for neutrino mass generation. A remarkable opportunity has emerged to search for such oscillations with the construction of the European Spallation Source (ESS). A collaboration has been formed which has proposed a search at the ESS, which would provide a sensitivity to the oscillation probability which is three orders of magnitude greater than that achieved at an ILL experiment at which the present best limit on free neutron-antineutron oscillations was obtained.
           }

\FullConference{The European Physical Society Conference on High Energy Physics\\
		22--29 July 2015\\
		Vienna, Austria}

\begin{document}

\section{Introduction}
The observation of baryon number violation (BNV) would address a number of open questions in modern physics. BNV is required to understand the matter-antimatter symmetry of the universe~\cite{Sakharov:1967dj}. Many models which explain  non-zero neutrino masses also prescribe BNV~\cite{Mohapatra:1980qe}. Even within the Standard Model (SM) baryon number is subject only to an approximate conservation law. At the perturbative level baryon number conservation arises due to the specific matter content in the SM, and corresponds to a so-called  ``accidental" symmetry. The SM predicts BNV to occur via rare non-perturbative electroweak instanton processes~\cite{Adler:1969gk,'tHooft:1976fv} (the quantum number $B-L$ is respected by the SM and not $B$ and $L$ separately).  Furthermore, precision tests of the Equivalence Principle~\cite{Adelberger:1990xq} offer no evidence for a long range force coupled to baryon number and thus a local gauge symmetry forbidding BNV. Consequently, BNV occurs as a generic feature of many proposed extensions to the SM~\cite{Barbier:2004ez}.  A promising means of searching for BNV is via the observation of the $\Delta B=2$ process, neutron-antineutron oscillation~\cite{Mohapatra:2009wp,Phillips:2014fgb}. In this paper, a proposed new experiment~\cite{EOInnbar} to look for such oscillations at the European Spallation Source (ESS) is outlined. The experiment would be sensitive to oscillation probabilities up to three orders of magnitude lower than has previously been obtained using free neutrons.


There exists a symbiosis between neutron-antineutron oscillations and neutrino physics via the quantum number $B-L$. A popular model explaining non-zero neutrino mass is the see-saw mechanism~\cite{see-saw}. In this approach neutrinos possess a Majorana component and lepton number is violated by two units. Evidence for $\Delta L=2$ processes are sought with, eg,  double neutrinoless beta decay searches~\cite{Elliott:2012sp}. Since $B-L$ (the true anomaly-free SM symmetry) is also violated by two units it would be natural to expect $\Delta B=2$ processes. In addition to the complementarity with neutrino physics, neutron-antineutron oscillation features in a number other models of new physics. Examples include $R$-parity violating supersymmetry~\cite{Barbier:2004ez} and post-sphaleron baryogesesis~\cite{Babu:2006xc}. Values of the BNV mass scale for which observable oscillations take place exceed those attainable at colliders.  Using a six-fermion BNV operator and dimensional reasoning mass scales of $10-1000$~TeV are obtained while other approaches (also leading to an observable signature) predict scales near the grand unified mass~\cite{Mohapatra:2009wp}.  A further motivation for searching for oscillations was recently provided by the observation that such processes violate not only baryon number but also $CP$~\cite{Berezhiani:2015uya}, thereby addressing two of the Sakharov conditions~\cite{Sakharov:1967dj} for baryogenesis.

Setting aside the substantial theoretical motivation,  a strictly experimentalist consideration of BNV hunting highlights the importance of neutron-antineutron oscillation searches.  In an oscillation experiment only the violation of baryon number is sought, and not that of other hitherto conserved quantities. Single nucleon decay searches (eg, $p\rightarrow \pi^0 e^+$) require lepton number violation ; among other reasons this ensures angular momentum conservation. Only searches for free neutron oscillation~\cite{Fidecaro:1985cm, BaldoCeolin:1994jz} and anomalous nuclear decays, under the neutron oscillation~\cite{nnbartrapped} or dinucleon decay-hypothesis~\cite{dinucleon}, offer high precision sensitivity to BNV-only processes. The most competitive limits for the free neutron oscillation time have hitherto been produced at ILL~\cite{BaldoCeolin:1994jz} ($\sim 3 \times 10^8$s) and Super-Kamiokande~\cite{Abe:2011ky} ($\sim 1 \times 10^8$s, after a correction for nuclear effects).

Of the class of experiments which search for BNV-only processes, free neutron oscillation searches possess both the cleanest experimental and theoretical environments in which to perform the search and quantify the results of the search. Owing to improvements in neutronics, particle identification technology and a longer running time, the proposed experiment at the ESS~\cite{EOInnbar} will have a sensitivity in oscillation probability which is to up to three orders of magnitude greater than at ILL.

This paper is organised as follows. Descriptions of the ESS and the neutron moderator are given. The plan for the transmission of neutrons to the detector is outlined, followed by a description of the detector. Finally, the collaboration which aims to conduct the experiment, as well a provisional timescale for the work, are briefly described.

\section{Overview of ESS and the proposed experiment}\label{sec:ess}
Currently under construction, the ESS is a multi-disciplinary research laboratory which will house the world's most powerful neutron source~\cite{ESS-TDR}. The ESS will comprise a 2.86 ms long proton pulse at 2 GeV energy at a repetition rate of 14 Hz which impacts on a rotating tungsten target. Spallation neutrons emerging from a system of moderators and reflectors are delivered to the beam ports and are then guided with a neutron supermirror to the instrument.  For the experiment described here, neutrons would be transported through in a vacuum through a magnetically shielded beam pipe over $200--300$m to a target with which anti-neutrons could annihilate. Magnetic shielding is necessary to suppress the energy split between neutron and antineutron states ($\Delta E=\bar{\mu} \cdot \bar{B}$) which would occur in a $B$-field due do the particles' dipole moments $\pm \bar{\mu}$, and which would inhibit the oscillation process. A detector surrounding the target would record the final states emerging from an annihilation as well as monitoring background processes.

The quantity of merit for a neutron-antineutron search is $N_n \cdot t^2$ where $N_n$ is the free neutron flux reaching the target and $t$ is the free flight time of the neutron.

For a high sensitivity (high $N_n \cdot t^2$) search the following criteria must be met:
\begin{itemize}
\item The moderator must deliver a beam of slow, cold neutrons (energy $< 5$meV) at high intensity, maximising $t$ and $N_n$, respectively.  An overall lower neutron spectrum emission also increases the transport efficiency of the supermirror neutron reflector.
\item The beam port must correspond to a large opening angle for neutron emission.
\item A long beamline to increase $t$.
\item Long running time.
\end{itemize}

A number of factors drive the improvement in sensitivity of the proposed experiment compared to the work at ILL. An important contribution to increased sensitivity is due to the use of a large elliptical focusing supermirror reflector which directs off-axis neutrons to the detector with a single reflection\footnote{Each reflection effectively ''resets the clock" prior to a putative neutron antineutron oscillation.}. Furthermore, a larger detector with enhanced particle identification is possible, as is a longer running time. 

\section{Moderator system}
\begin{figure}[tb]
  \setlength{\unitlength}{1mm}
  \begin{center}
\includegraphics[width=0.80\linewidth, angle=90]{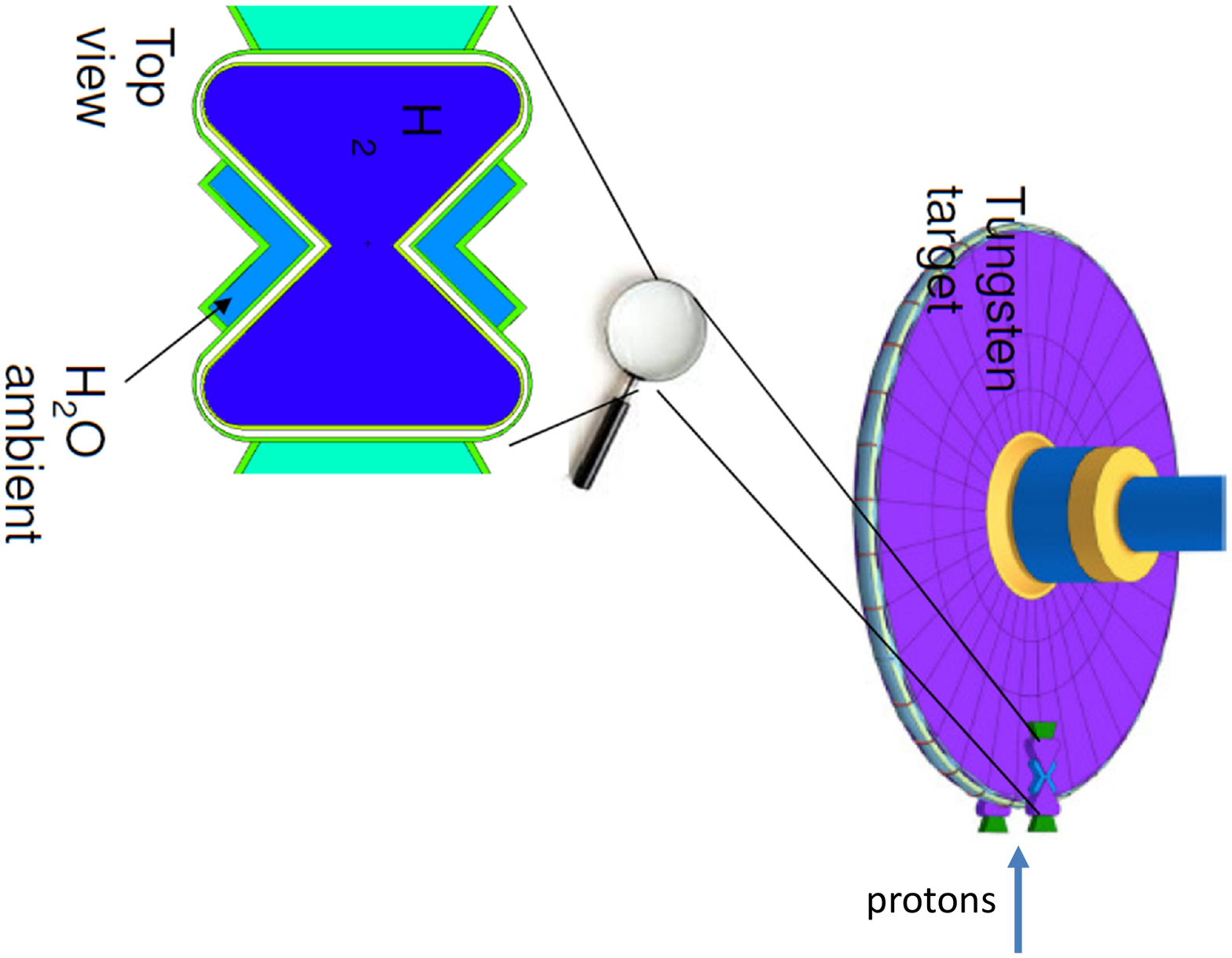}
    \end{center}
    \vspace{-2.25cm}
  \caption{Overview of the ESS and the moderator system.
  }
  \label{fig:wheelandmoderator}
\end{figure}

A conceptual overview of the ESS and moderator system is given Figure~\ref{fig:wheelandmoderator}. A ''butterfly" moderator design has been chosen which comprises cold regions of para-$H_2$ at around $20$~K, and water at ambient temperature . The upper (lower) moderator has a height of 3cm (6cm). This represents the most optimal choice for the brightness of the ESS. In addition, it provides a flexibility such that the placing of the instruments can be chosen according to which moderator is most appropriate for the experiment. For the work proposed here, a beam port is available such that neutrons from both moderators can be exploited.

\begin{figure}[tb]
  \setlength{\unitlength}{1mm}
  \begin{center}
\includegraphics[width=1.00\linewidth, angle=-90]{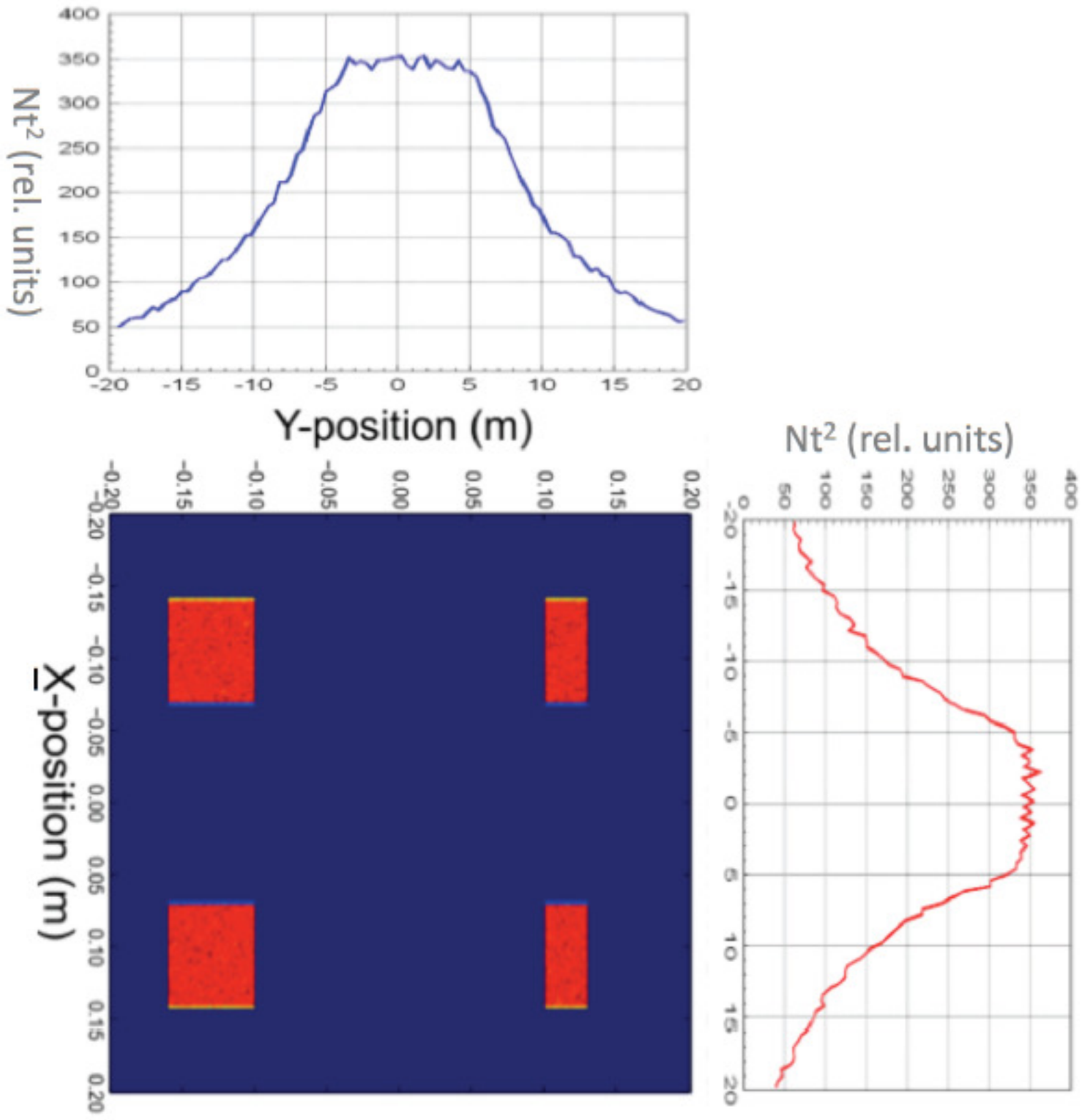}
    \end{center}
    \vspace{-2.75cm}
  \caption{ View of the butterfly moderators from the supermirror showing the cold moderators in red. The top and left figures show the relative guiding efficiency by a mirror centred on the middle point as a function of the point of emission.
  }
  \label{fig:neutronics}
\end{figure}

Figure~\ref{fig:neutronics} shows a cross sectional view of the moderator system from the suppermirror. The cold moderators are shown in red.  Here, the $x$ and $y$ coordinates are axes map out the horizontal plane. The figures to the top and left  show the relative guiding efficiency by a truncated focusing ellipsoid mirror centred on the middle point as a function of the point of emission. The transport efficiency for neutrons produced in the cold region is $\sim 10$\% of that expected for neutrons produced at the ellipsoid's focal point. This drop in efficiencies near the cold region illustrates the need for a sophisticated mirror configuration which fully takes into account the design of the moderator. At present a ''clover" assembly of four quarter ellipsoids, each one centred on a moderator, is being considered, as is a simpler design comprising an ellipsoid focused on aa cold moderator.

\begin{figure}[tb]
  \setlength{\unitlength}{1mm}
  \begin{center}
\includegraphics[width=0.80\linewidth, angle=0]{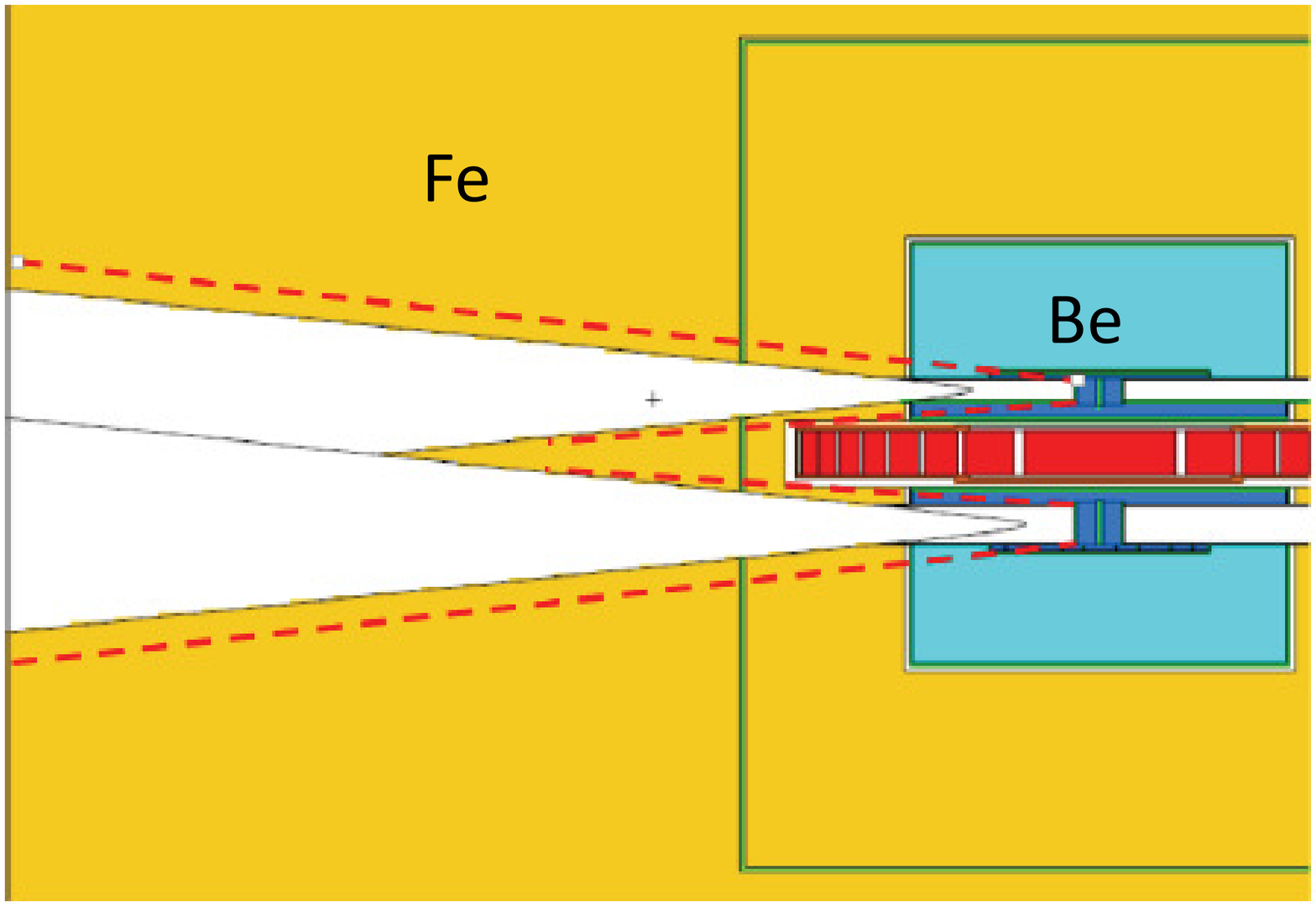}
    \end{center}
    \vspace{-.75cm}
  \caption{ Nominal (white region to the left of the neutron source) and enlarged (region enclosed by dashed lines) conical penetration through the Be reflector and Fe shield).
  }
  \label{fig:moderatorandreflector}
\end{figure}

An increase to the nominal angular acceptance for produced neutrons is another area in which sensitivity can be enhanced. As seen in Figure~\ref{fig:moderatorandreflector} losses will occur due to the presence of the Fe shield and Be reflector system. Figure~\ref{fig:moderatorandreflector} shows a possible adjustment to the design of the beam port for the proposed work. Here, the parts of the shield and reflector system would be removed to allow a greater conical penetration corresponding to an increase in sensitivity of $\sim 2$. Inserts could be made such that the full system is restored for experiments after the proposed work. The study of such a scheme including its on the other experiments is underway.

\begin{figure}[tb]
  \setlength{\unitlength}{1mm}
  \begin{center}
\includegraphics[width=0.80\linewidth, angle=0]{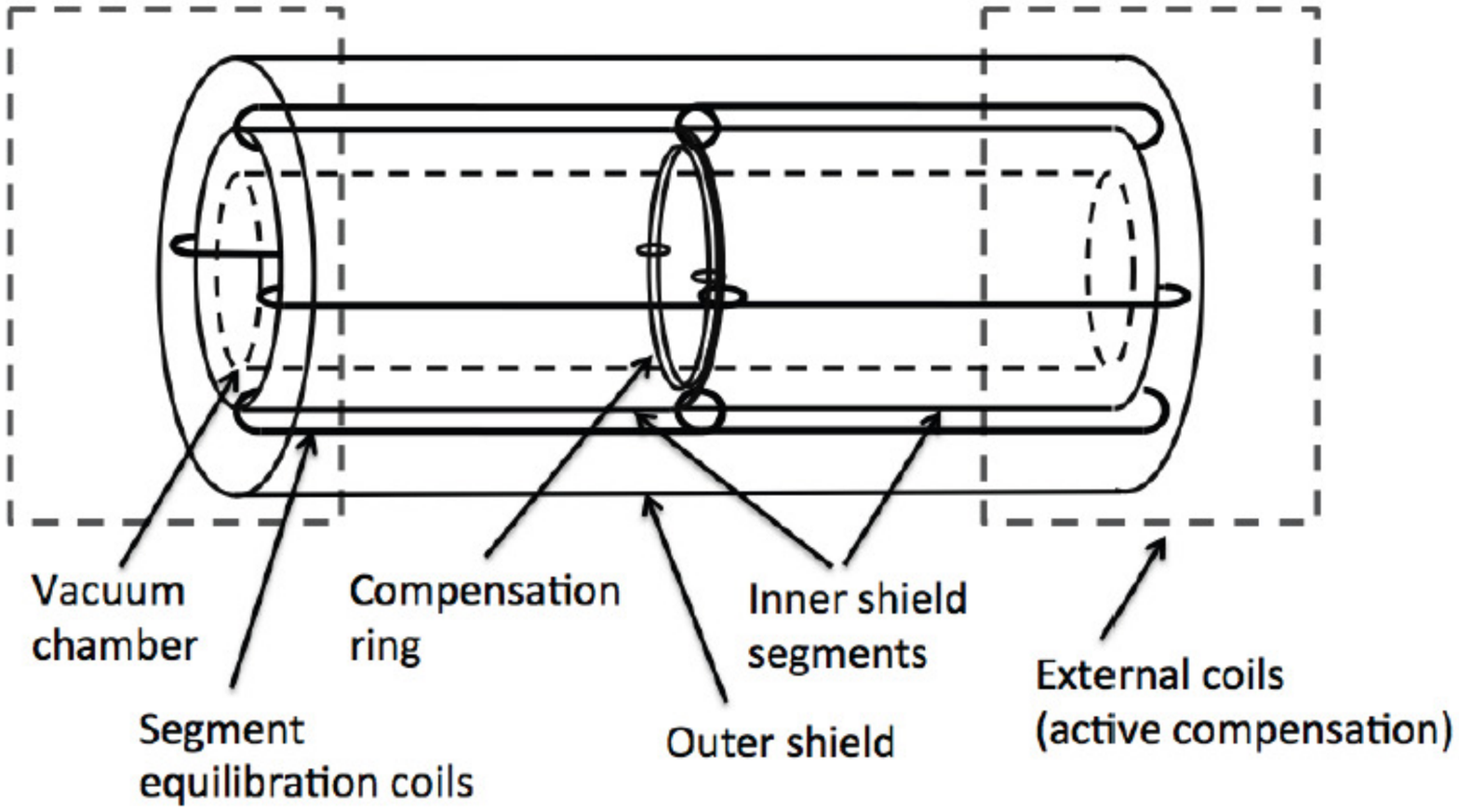}
    \end{center}
    \vspace{-1.75cm}
  \caption{ Schematic overview of the planned shielding.
  }
  \label{fig:shield}
\end{figure}

\section{Magnetics}
As mentioned in Section~\ref{sec:ess}, the neutrons must be transported in a magnetically shielded vaccum. For the proposed work this corresponds to a vacuum of $10^{-5}$ mbar and a magnetic field of less than 5nT along the neutron flight path.

The target vaccum can be achieved with a vacuum chamber comprising highly non-magnetic materials, eg Al, with turbo molecular pumps, mounted outside of the magnetically shielded area. Magnetic fields of  less than 5 nT have been achieved over large volumes (see, for example, Ref.~\cite{m1}). For the planned experiment, a shielding concept will be used based on an aluminium vacuum chamber, a two layer passive shield made from magnetizable alloy for transverse shielding, and end sections made from passive and active components for longitudinal shielding, as shown in Figure~\ref{fig:shield}.

\section{Detector}
\begin{figure}[tb]
  \setlength{\unitlength}{1mm}
  \begin{center}
\includegraphics[width=0.80\linewidth, angle=90]{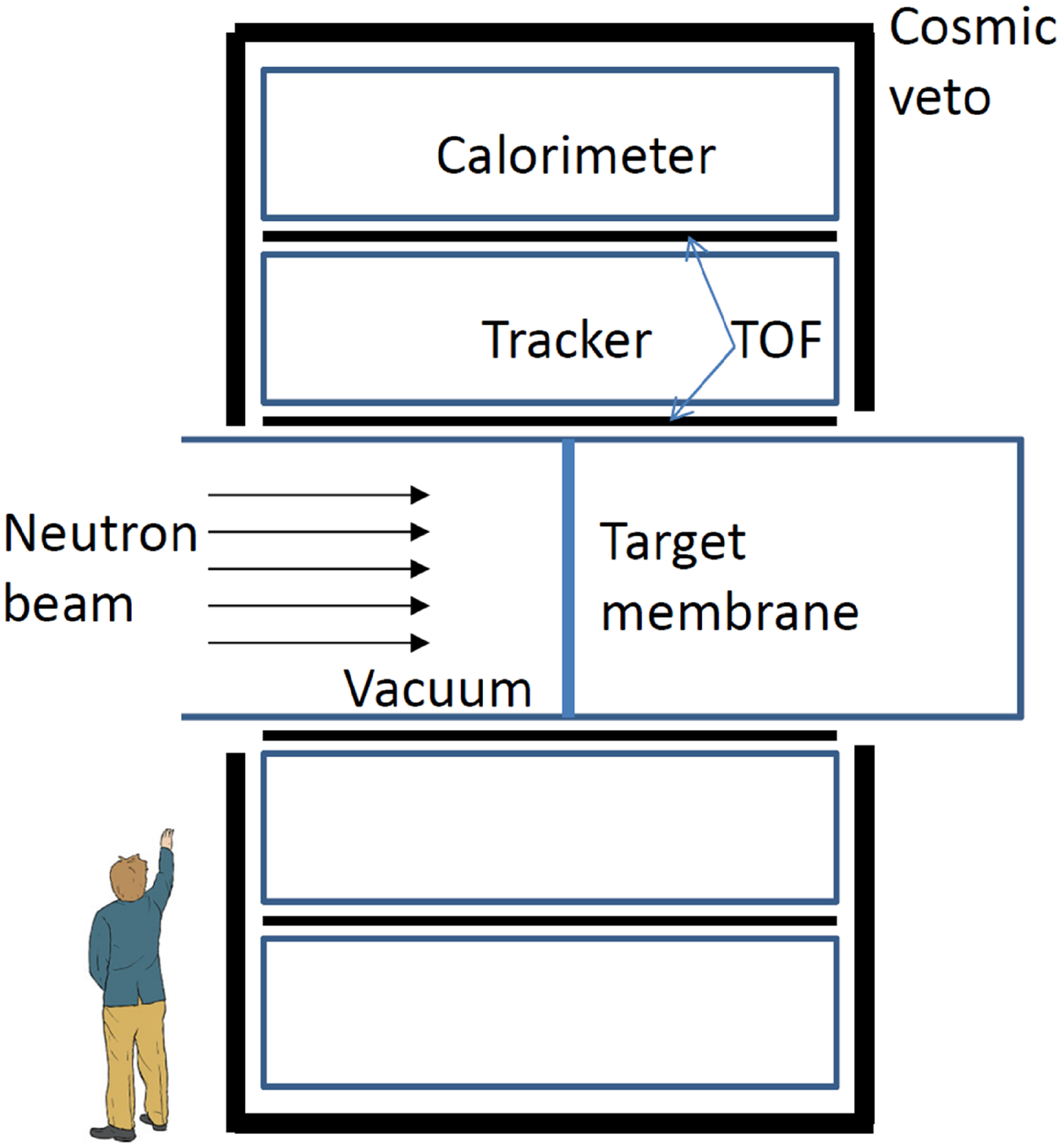}
    \end{center}
    \vspace{-2.15cm}
  \caption{Schematic overview of the detector.
  }
  \label{fig:detector}
\end{figure}

Detector design is guided by the need for high efficiency for antineutron detection and the ability to maintain a low background yield. A typical annihilation signature would be a multi-pion final state. A schematic diagram of the detector is given in Figure~\ref{fig:detector}. The following components are envisaged and studies on the various technology options underway:

\begin{itemize}
\item A annihilation target. One option is a $^{12}C$ disk of diameter $1$~m and which is 100~$\mu$m thick. An alternative is a target made of $^{10}Be$ which may have better potential to capture photons from background processes. A two target system is also being considered; an antineutron would annihilate in the first target so a second target could be used for background monitoring.
\item A charged particle tracker, necessary for the determination of pion momenta and the vertex position. Different technologies are under consideration, for a straw tube-based drift chamber or a Time Projection Chamber. However, any tracking system will need at least some layers with fast readout (i.e. straws) to allow tracking information to be included in the trigger.
\item A calorimeter must accurately measure photon and pion energies in order to reconstruct the final state invariant mass. Depending on the technology choice, precision timing information from the calorimeter or a calorimeter+time-of-flight configuration would be available. This is necessary for establishing the time of an annihilation event, and for rejecting false vertex reconstructions due to cosmic ray showers. The calorimeter will also need to handle high pile-up rates from gamma production at the target.
\item A trigger exploiting all read-out channels enabling a highly selective system to collect signal and background candidate events.
\item A dedicated cosmic veto system to reject background.
\end{itemize}

\section{Collaboration and time-scales}
A growing collaboration has been formed with the aim of carrying out the proposed work. Working groups corresponding to specific aspects of the experiment have been established and a number of workshops have taken place. An expression of interest (EOI) with signatories from 26 institutes and 8 countries has been sent~\cite{EOInnbar}.

A provisional time-scale consists of the preparation a Technical Design Report in 2017, construction in 2019, commissioning in 2022, and data-taking ''physics" runs in 2023-2025.

\section{Summary}
Strong theoretical motivations addressing open questions in modern physics such as the matter-antimatter asymmetry and the possible majorana nature of the neutrino imply the existence of BNV processes. A promising means of searching for BNV is via the neutron-antineutron oscillation signature. A new high precision search is being planned for neutron-antineutron oscillation processes at the ESS. The sensitivity to the oscillation probability would be around three orders of magnitude compared with the last such experiment. A collaboration to carry has been formed with the aim of performing the experiment.

\end{document}